\pdfoutput=1

\documentclass[11pt,twoside,a4paper,cmspaper,final,collab]{cms-tdr}
\def\svnVersion{81c7d55-D}\def\svnDate{2019/08/14}\def\cmsCernNoTag{CERN-EP-2019-110}\def\cmsCernDate{\today}\def\cmsMessage{"Published in Physics Letters B as \href{http://dx.doi.org/10.1016/j.physletb.2019.134952}{\doi{10.1016/j.physletb.2019.134952}.}"}

\begin{document}\cmsNoteHeader{B2G-18-006}

\hyphenation{had-ron-i-za-tion}
\hyphenation{cal-or-i-me-ter}
\hyphenation{de-vices}
\RCS$HeadURL$
\RCS$Id$

\newlength\cmsFigWidth
\newlength\cmsTabSkip\setlength{\cmsTabSkip}{1ex}
\ifthenelse{\boolean{cms@external}}{\setlength\cmsFigWidth{0.49\textwidth}}{\setlength\cmsFigWidth{0.65\textwidth}}
\ifthenelse{\boolean{cms@external}}{\providecommand{\cmsLeft}{upper\xspace}}{\providecommand{\cmsLeft}{left\xspace}}
\ifthenelse{\boolean{cms@external}}{\providecommand{\cmsRight}{lower\xspace}}{\providecommand{\cmsRight}{right\xspace}}
\ifthenelse{\boolean{cms@external}}{\providecommand{\cmsTable}[1]{#1}}{\providecommand{\cmsTable}[1]{\resizebox{\textwidth}{!}{#1}}}

\newcommand{\X}{{\HepParticle{X}{}{}}\xspace}
\newcommand{\G}{{\HepParticle{G}{}{}}\xspace}
\newcommand{\V}{{\HepParticle{V}{}{}}\xspace}
\newcommand{\PVpr}{{\HepParticle{V}{}{\prime}}\xspace}
\newcommand{\PWprpm}{{\HepParticle{W}{}{\prime\pm}}\xspace}
\newcommand{\VV}{\ensuremath{\V\V}\xspace}
\newcommand{\VH}{\ensuremath{\V\PH}\xspace}
\newcommand{\HH}{\ensuremath{\PH\PH}\xspace}
\newcommand{\mX}{\ensuremath{m_{\X}}\xspace}
\newcommand{\mVpr}{\ensuremath{m_{\PVpr}}\xspace}
\newcommand{\mj}{\ensuremath{m_{\mathrm{j}}}\xspace}
\newcommand{\mjj}{\ensuremath{m_{\mathrm{jj}}}\xspace}
\newcommand{\nsub}{\ensuremath{\tau_{21}}\xspace}
\newcommand{\gV}{\ensuremath{g_\text{V}}\xspace}
\newcommand{\cH}{\ensuremath{c_\text{H}}\xspace}
\newcommand{\cF}{\ensuremath{c_\text{F}}\xspace}
\newcommand{\gH}{\ensuremath{g_\text{H}}\xspace}
\newcommand{\gF}{\ensuremath{g_\text{f}}\xspace}
\newcommand{\ktilde}{\ensuremath{\widetilde{\kappa}}\xspace}
\newcommand{\SI}{s}
\newcommand{\BA}{b}
\newcommand{\SB}{s, b}
\newcommand{\FP}{f}
\newcommand{\TE}{t}
\newcommand{\TB}{t, b}
\newcommand{\NO}{no}
\newcommand{\YES}{yes}

\cmsNoteHeader{B2G-18-006}
\title{Combination of CMS searches for heavy resonances decaying to pairs of bosons or leptons}
\date{\today}

\abstract{
A statistical combination of searches for heavy resonances decaying to pairs of bosons or leptons is presented. The data correspond to an integrated luminosity of 35.9\fbinv collected during 2016 by the CMS experiment at the LHC in proton-proton collisions at a center-of-mass energy of 13\TeV. The data are found to be consistent with expectations from the standard model background. Exclusion limits are set in the context of models of spin-1 heavy vector triplets and of spin-2 bulk gravitons. For mass-degenerate \PWpr and \PZpr resonances that predominantly couple to the standard model gauge bosons, the mass exclusion at 95\% confidence level of heavy vector bosons is extended to 4.5\TeV as compared to 3.8\TeV determined from the best individual channel. This excluded mass increases to 5.0\TeV if the resonances couple predominantly to fermions.
}

\hypersetup{
pdfauthor={CMS Collaboration},
pdftitle={Combination of CMS searches for heavy resonances decaying to pairs of bosons or leptons},
pdfsubject={CMS},
pdfkeywords={CMS, B2G, diboson, dilepton, HVT, graviton}
}

\maketitle

\section{Introduction}

Over the past half century, successive searches for heavy resonances in two-body decays have often led to discoveries of new states. At the CERN LHC, searches for \PWpr and \PZpr resonances (collectively referred to as \PVpr) that couple through the electroweak (EW) interaction to standard model (SM) particles have been performed by ATLAS and CMS in final states with two SM bosons~\cite{Sirunyan:2017acf,Sirunyan:2018ivv,Sirunyan:2018iff,Sirunyan:2018hsl,Sirunyan:2017jtu,Sirunyan:2017wto,Sirunyan:2018qob,Sirunyan:2018fuh,Sirunyan:2017isc,Sirunyan:2018qca,Aaboud:2017eta,Aaboud:2017fgj,Aaboud:2017itg,Aaboud:2017gsl,Aaboud:2018ohp,Aaboud:2017rel,Aaboud:2017ahz,Aaboud:2017cxo}, two leptons~\cite{Sirunyan:2018mpc,Sirunyan:2018exx,Aaboud:2017efa,Aaboud:2017buh}, and two light-flavor~\cite{Sirunyan:2018xlo,Aaboud:2017yvp} or heavy-flavor quarks~\cite{Sirunyan:2018ryr,Sirunyan:2017vkm,Aaboud:2018tqo,Aaboud:2018mjh,Aaboud:2018juj}.
These new states would couple predominantly to either SM fermions, as in the case of minimal \PWpr and \PZpr models~\cite{Grojean:2011vu,Barger:1980ix,1126-6708-2009-11-068}, or SM bosons, as in strongly coupled composite Higgs and little Higgs models~\cite{Contino2011,Marzocca2012,Bellazzini:2014yua,Lane:2016kvg,Han:2003wu,Schmaltz,Perelstein2007247}. In addition, models based on warped extra dimensions provide a candidate for a massive resonance, such as the spin-2 first Kaluza--Klein excitation of the graviton (\G)~\cite{Randall:1999ee,Randall:1999vf,Agashe:2007zd}. In bulk graviton models~\cite{Fitzpatrick:2007qr}, the SM fermions and gauge bosons are located in the bulk five-dimensional spacetime, and the graviton has a sizable branching fraction to pairs of \PW, \PZ, and \PH bosons.

This Letter describes a statistical combination of CMS searches for heavy resonances that decay to pairs of bosons or leptons~\cite{Sirunyan:2017jtu,Sirunyan:2017isc,Sirunyan:2018iff,Sirunyan:2017acf,Sirunyan:2017wto,Sirunyan:2018qob,Sirunyan:2018ivv,Sirunyan:2018fuh,Sirunyan:2018hsl,Sirunyan:2018qca,Sirunyan:2018mpc,Sirunyan:2018exx}. The event selection, the simulated samples, the background estimation, and the systematic uncertainties of the individual analyses are unchanged. The results are used to set exclusion limits on models that invoke heavy vector resonances and on a model with a bulk graviton.

The SM bosons produced in the decay of resonances \X with masses \mX that exceed 1\TeV are expected to have a large Lorentz boost~\cite{Dorigo:2018cbl}. The decay products of the SM bosons are therefore highly collimated, requiring dedicated techniques for their identification and reconstruction. In the case of hadronic decays, the pair of quarks produce a single large-cone jet with a two-prong structure. In addition, Higgs bosons can also be identified by tagging the {\cPqb} quarks originating from their decays.
In models where heavy vector bosons couple predominantly to fermions, the combined contribution from the leptonic decays $\PWpr\to\ell\nu$ and $\PZpr\to\ell\ell$ dominates, and these channels~\cite{Sirunyan:2018mpc,Sirunyan:2018exx} are included in the combination, providing complementary sensitivity to the searches in diboson channels. The analyses considered in this Letter are based on a data sample of proton-proton ($\Pp\Pp$) collisions at a center-of-mass energy of $13\TeV$ collected by the CMS experiment in 2016, and corresponding to an integrated luminosity of 35.9\fbinv. A similar combination performed on a comparable set of data has been recently published by ATLAS~\cite{Aaboud:2018bun}.

\section{The CMS detector and event reconstruction}\label{sec:detector}

The CMS detector features a silicon pixel and strip tracker, a lead tungstate crystal electromagnetic calorimeter (ECAL), and a brass and scintillator hadron calorimeter (HCAL), each composed of a barrel and two endcap sections. These detectors reside within a superconducting solenoid, which provides a magnetic field of 3.8\unit{T}. Forward calorimeters extend the pseudorapidity coverage up to $\abs{\eta} < 5.2$. Muons are measured in gas-ionization detectors embedded in the steel flux-return yoke outside the solenoid.
A detailed description of the CMS detector, together with a definition of the coordinate system and the kinematic variables, can be found in Ref.~\cite{Chatrchyan:2008zzk}.

The information from various elements of the CMS detector is used by a particle-flow (PF) algorithm~\cite{Sirunyan:2017ulk} to identify stable particles reconstructed in the detector as electrons, muons, photons, and charged or neutral hadrons. The energy of electrons is determined from a combination of the electron momentum, as determined in the tracker, the energy of the corresponding ECAL cluster, and the energy sum of all bremsstrahlung photons spatially compatible with originating from the electron track. The dielectron mass resolution for $\PZ \to \Pe \Pe$ decays ranges from 1.9\% when both electrons are in the ECAL barrel, to 2.9\% when both electrons are in the endcaps~\cite{Khachatryan:2015hwa}. The momentum of muons is obtained from the curvature of the corresponding track. The \pt resolution in the barrel is better than 7\% for muons with \pt up to 1\TeV~\cite{Sirunyan:2018mu}. The $\tau$ leptons that decay to hadrons and a neutrino (labeled $\tauh$) are reconstructed by combining one or three charged PF candidates with up to two neutral pion candidates~\cite{Khachatryan:2015dfa}. The $\Pe$, $\mu$, and $\tauh$ with large momenta are required to be isolated from other hadronic activity in the event. The energy of charged hadrons is determined from a combination of their momenta measured in the tracker and the matching ECAL and HCAL energy deposits. Finally, the energy of neutral hadrons is obtained from the corresponding corrected ECAL and HCAL energies.

Jets are reconstructed from PF candidates clustered with the anti-\kt algorithm~\cite{Cacciari:2008gp}, with a distance parameter 0.4 (AK4 jets) or 0.8 (AK8 jets), using the {\FASTJET} 3.0 package~\cite{Cacciari:2011ma,Cacciari:2008gn}. The geometrical overlaps between jets and isolated leptons, and between AK4 jets and the AK8 jet, are handled by assigning the overlapping four-momentum to the lepton, and the AK8 jet, respectively.
The four-momenta of the AK4 and AK8 jets are obtained by clustering candidates passing the charged-hadron subtraction algorithm~\cite{CMS-PAS-JME-14-001}. The contribution of neutral particles originating from additional $\Pp\Pp$ interactions within the same or neighboring bunch crossings (pileup) is proportional to the jet area, and is estimated using the median area method implemented in {\FASTJET}, and then subtracted from the jet energy. The jet energy resolution, after the application of corrections to the jet energy, is 4\% at 1\TeV~\cite{Khachatryan:2016kdb}.
The missing transverse momentum vector \ptvecmiss is computed as the negative vector sum of the transverse momenta of all the PF candidates in an event, and its magnitude is denoted as \ptmiss~\cite{CMS-PAS-JME-17-001}. The \ptvecmiss is corrected for adjustments to the energy scale of the reconstructed AK4 jets in the event.

While AK4 jets are used for single quarks, AK8 jets are adopted to reconstruct large momentum SM bosons that decay to quarks.
Reconstructing the AK8 jet mass (\mj) and substructure relies on the pileup-per-particle identification algorithm~\cite{Bertolini2014,CMS-PAS-JME-14-001}. The contributions from soft radiation and additional interactions are removed using the soft-drop algorithm~\cite{Dasgupta:2013ihk,Larkoski:2014wba}, with parameters $\beta = 0$ and $z_\text{cut} = 0.1$.
Dedicated mass corrections, obtained from simulation and data in a region enriched with $\ttbar$ events with merged $\PW(\cPq\bar{\cPq}')$ decays, are applied to the jet mass to reduce any residual jet \pt dependence~\cite{CMS-PAS-JME-16-003,Sirunyan:2017wto}, and to match the jet mass scale and resolution observed in data. The measured soft-drop jet mass resolution is approximately constant at 10\% in the considered range of jet \pt~\cite{CMS-PAS-JME-16-003}.
Exclusive \mj intervals labeled $m_{\PW}$, $m_{\PZ}$, and $m_{\PH}$, which range from 65 to 85, 85 to 105, and 105 to 135\GeV, respectively, are defined according to the SM boson masses and the jet mass resolution.

Hadronic decays of $\PW$ and $\PZ$ bosons are identified using the ratio between 2-subjettiness and 1-subjettiness~\cite{Thaler:2010tr}, $\nsub=\tau_{2}/\tau_{1}$. The variables \mj and \nsub are calibrated using a top quark-antiquark sample enriched in hadronically decaying $\PW$ bosons~\cite{Khachatryan:2014vla}.
The decay of a Higgs boson to a pair of {\cPqb} quarks is identified using one of two {\cPqb} tagging algorithms, depending on the background composition. The first consists of a dedicated {\cPqb} tagging discriminator, specifically designed to identify a pair of {\cPqb} quarks clustered in a single jet~\cite{Sirunyan:2017ezt}. The second relies on splitting the AK8 jet into two subjets, then applying the combined secondary vertex algorithm~\cite{Sirunyan:2017ezt} to the subjets. The latter is also applied to AK4 jets to identify isolated {\cPqb} quarks in the event.

\section{Signal modeling}\label{sec:mcsimulation}

The response of the CMS detector to the production and decay of heavy resonances is evaluated through simulated events, which are reconstructed using the same algorithms as used in collision data. The spin-1 gauge bosons, \PWpr and \PZpr, are simulated at leading order (LO) using the \MGvATNLO 2.4.2 matrix element generator~\cite{MadGraph}, in the heavy vector triplet (HVT) framework~\cite{deBlas:2012qp,Pappadopulo2014}, which introduces a triplet of heavy vector bosons with similar mass, of which one is neutral ($\PZpr$) and two are electrically charged ($\PWprpm$). The coupling strength of the heavy vector bosons to SM bosons and fermions is determined from the combinations $\gH = \gV\cH$ and $\gF = g^2 \cF / \gV$, respectively. The parameter $\gV$ is the strength of the new interaction; $\cH$ characterizes the interaction between the HVT bosons, the Higgs boson, and longitudinally polarized SM vector bosons; $\cF$ represents the direct interaction between the $\PVpr$ bosons and the SM fermions; $g$ is the SM $\mathrm{SU(2)_L}$ gauge coupling constant.
The HVT framework is presented in two scenarios, henceforth referred to as model~A and model~B, depending on the couplings to the SM particles~\cite{Pappadopulo2014}. In the former, the coupling strengths to the SM bosons and fermions are comparable, and the new particles decay primarily to fermions. In the latter, the couplings to the SM fermions are small, and the branching fraction to the SM bosons is nearly 100\%. Events are simulated with different \mX hypotheses from 800 to 4500\GeV for \X decays to bosons, and 800 to 5500\GeV for \X decays to leptons, assuming a negligible resonance width compared to the experimental resolution (3--20\%). The kinematic distributions of the signal do not depend on the choice of benchmark scenario, and the samples are reweighted according only to the cross sections and branching fractions.

The bulk graviton events are also simulated at LO using the same \MGvATNLO generator. The cross section and the width of the bulk graviton mainly depend on its mass and the ratio $\ktilde\equiv \kappa/\overline{M}_\text{Pl}$, where $\kappa$ is a curvature factor of the model and $\overline{M}_\text{Pl}$ is the reduced Planck mass~\cite{Fitzpatrick:2007qr,Goldberger:1999uk}. The graviton signals are generated assuming $\ktilde = 0.5$, which guarantees that the width of the graviton is smaller than the experimental resolution.

The signal events are generated using the NNPDF 3.0~\cite{Ball:2014uwa} parton distribution functions (PDFs), and are interfaced to~\PYTHIA~8.205~\cite{Sjostrand:2014zea} for parton showering and hadronization, adopting the MLM matching scheme~\cite{Alwall:2007fs}.
Simulated pileup interactions are superimposed on the hard process, and their frequency distribution is weighted to match the number of interactions per bunch crossing observed in data. Generated events are processed through the CMS detector simulation, based on {\GEANTfour}~\cite{Agostinelli:2002hh}.

\section{Search channels}\label{sec:analyses}

The search strategies for the analyses contributing to the combination are summarized in this Section.
More details are provided in the relevant publications~\cite{Sirunyan:2017jtu,Sirunyan:2017isc,Sirunyan:2018iff,Sirunyan:2017acf,Sirunyan:2017wto,Sirunyan:2018qob,Sirunyan:2018ivv,Sirunyan:2018fuh,Sirunyan:2018hsl,Sirunyan:2018qca,Sirunyan:2018mpc,Sirunyan:2018exx}.

\subsection{Fully hadronic diboson channels}\label{ssec:hadronic}

Diboson resonances have been searched for in several final states, depending on the decay modes of the bosons. The final states targeted were $\VV\to\qqbar\qqbar$~\cite{Sirunyan:2017acf}, $\VH\to\qqbar\bbbar$~\cite{Sirunyan:2017wto}, and $\HH\to\bbbar\bbbar$ final states~\cite{Sirunyan:2017isc,Sirunyan:2018qca}. Each boson is reconstructed as a two-prong, large-cone jet, so that for diboson resonances, the two jets would recoil against each other. The presence of a diboson resonance would be observed in the dijet invariant mass spectrum \mjj. The \PW~and \PZ~bosons are identified (\V~tagged) via the \mj and \nsub variables, and {\cPqb} tagging is used to identify {\cPqb} quarks from Higgs bosons in addition to \mj. Although the signal yield is large, because of the large fraction of Higgs boson decays to {\cPqb} quarks, these channels are subject to an overwhelming background from quantum chromodynamics (QCD) multijet production.

In the \VV and \VH analyses, the background is estimated directly from data, assuming that the invariant mass distribution of the background can be described by a smooth, parameterizable, monotonically decreasing function of \mjj. The signal template, based on a Gaussian core, is fitted to the data simultaneously with the background function.
The \HH analyses also use an additional region in the fit, obtained by inverting the {\cPqb} tagging selection on the \PH candidates, which constrains the parameters of the background function.

\subsection{Semi-leptonic diboson channels}\label{ssec:semilept}

Searches for \VV, \VH, and \HH resonances have been performed in channels where one of the SM bosons decays to leptons and the other to quarks ($\PZ\V\to\nu\nu\qqbar$~\cite{Sirunyan:2018ivv}, $\PW\V\to\ell\nu\qqbar$~\cite{Sirunyan:2018iff}, $\PZ\V\to\ell\ell\qqbar$~\cite{Sirunyan:2018hsl}, $\PZ\PH\to\nu\nu\bbbar$~\cite{Sirunyan:2018qob}, $\PW\PH\to\ell\nu\bbbar$~\cite{Sirunyan:2018qob}, $\PZ\PH\to\ell\ell\bbbar$~\cite{Sirunyan:2018qob}, $\V\PH\to\qqbar\tau\tau$~\cite{Sirunyan:2018fuh} and $\PH\PH\to\bbbar\tau\tau$~\cite{Sirunyan:2018fuh}). These final states represent an attractive alternative to all-jet final states, thanks to the large selection efficiencies and natural discrimination against multijet background stemming from the presence in the signal of energetic and isolated leptons or neutrinos.

The decay of a \PZ boson to neutrinos can be identified through its large \ptmiss, and the resonance mass can be inferred from the transverse mass between the \ptvecmiss and the jet originating from the hadronic decay of the other boson.
For the $\PW\to\ell\nu$ decay, there is a single, isolated lepton associated with a moderate \ptmiss, and the vector boson can therefore be reconstructed by imposing a constraint from the \PW~boson mass to recover the longitudinal momentum of the neutrino. In $\PZ\to\ell\ell$ decays, two opposite-sign, same-flavor leptons with a combined invariant mass compatible with the \PZ boson mass are used to determine accurately the \PZ boson four-momentum. A Higgs boson decaying to $\tau$ leptons is identified from dedicated $\tauh$ decay reconstruction and isolation techniques~\cite{Sirunyan:2018fuh}, and its mass is estimated through the measured momenta of the visible decay products of the $\tau$ leptons and the \ptvecmiss, with the {\sc SVfit} algorithm~\cite{Bianchini:2016yrt}. The boson that decays to a pair of quarks is reconstructed as an AK8 jet. The AK8 jet mass is required to be compatible with either the \PW and \PZ boson masses, denoted as $m_{\V}$ ($65 < \mj < 105\GeV$), or the Higgs boson mass ($105 < \mj < 135\GeV$).

In the semi-leptonic analyses, the main $\V$+jets background is estimated from a fit to data in the \mj sidebands of the hadronic jet, and extrapolated to the signal region using a transfer function (``$\alpha$ function'') obtained from simulation. The top quark pair production is estimated from simulation, but its normalization is rescaled to match the data in control regions obtained by requiring an additional {\cPqb}-tagged AK4 jet in the event~\cite{Sirunyan:2018qob,Sirunyan:2018ivv,Sirunyan:2018fuh}.

In addition, the $\PW\V\to\ell\nu\qqbar$ analysis introduces a novel signal extraction method based on a two-dimensional (2D) fit to data~\cite{Sirunyan:2018iff}. The backgrounds are separated into non-resonant and resonant categories depending on the presence or absence of \PW~bosons and top quarks in the jet mass spectrum, and are fitted simultaneously in the space of \mj and $m_{\PW\V}$, accounting for the correlation between the two variables.

\subsection{Fully leptonic diboson channels}\label{ssec:fullylep}

Searches for diboson resonances decaying to a pair of \PZ bosons have been performed in fully leptonic final states, with one boson undergoing the decay $\PZ\to\ell\ell$ and the other $\PZ\to\nu\nu$~\cite{Sirunyan:2017jtu}. The presence of the leptons and neutrinos defines a very clean final state with reduced backgrounds, but the small branching fraction makes this channel competitive only for small resonant mass values.

\subsection{Decays to a pair of fermions}\label{ssec:fermion}

The decay of a heavy resonance to a pair of fermions can be sizable when the couplings to SM fermions are large. If the resonance is electrically charged, as in the case of $\PWprpm$, the decay to a neutrino and an electron or muon yields a broad excess in the $\ell\nu$ transverse mass spectrum~\cite{Sirunyan:2018mpc}. If the new state is neutral, as in the case of \PZpr, a narrow resonance would emerge from the dielectron or dimuon ($\ell\ell$) invariant mass spectra~\cite{Sirunyan:2018exx}. The analyses of these fermionic decays extend to masses above 5\TeV, and employ selection techniques optimized to identify and measure very energetic electrons and muons in the detector~\cite{Sirunyan:2018mpc,Sirunyan:2018exx}.

\section{Event selection}\label{sec:selection}

The search regions of the analyses entering in the combination are statistically independent because of mutually exclusive selections on the number of leptons and their flavor, number of AK8 jets, and jet mass intervals. Analyses with hadronic final states reject events with isolated leptons or with large \ptmiss. Overlaps between channels that share the same lepton multiplicity are avoided by selecting different jet mass ranges. The $\PWpr\to\ell\nu$ search does not share any events with the $\PWpr\to\V\PW$ and $\PWpr\to\PW\PH$ analyses because of the requirements on the angular separation $\Delta\phi(\ell,\ptvecmiss)$ between the \ptvecmiss and the lepton direction. The $\PZpr\to\ell\ell$ analysis includes events with a dilepton invariant mass $m_{\ell\ell} > 120\GeV$, which is incompatible with the $70 < m_{\ell\ell} < 110\GeV$ selection used in the diboson channels, in which the \PZ boson is on-shell. The two searches for resonant $\PH\PH$ bosons that decay to {\cPqb} quarks have common events explicitly removed~\cite{Sirunyan:2018qca}.

In the $\PW\V\to\ell\nu\qqbar$ channel~\cite{Sirunyan:2018iff} the background is estimated using a 2D fitting technique that scans the full jet mass range, and is therefore not independent of the $\PW\PH\to\ell\nu\bbbar$ channel. For this reason, in the \PWpr, \PZpr, and \PVpr~interpretations, where the two signals might be present simultaneously, the ``$\alpha$ function'' is used to estimate the background instead. This method considers only events in the \mj regions of the \PW~and \PZ bosons, thereby preventing double counting of events in the Higgs boson mass region. The results of the alternative background estimation method are consistent with those obtained in the 2D fit, but the method is about 10\% less sensitive. The main selections that define the exclusivity of the analyses are summarized in Table~\ref{tab:analyses}.

\begin{table*}[!htb]
\centering
\topcaption{Summary of the main selections that guarantee the exclusivity between individual final states. The symbol $\ell$ represents an electron or a muon; $\tau$ leptons are considered separately. If an entry is not preceded by ``$\geq$'', the exact value is required. The AK4 {\cPqb} jets are additional {\cPqb} tagged AK4 jets that do not geometrically overlap with AK8 jets. The symbol ``\NA'' implies that no selection is applied.}
  \resizebox{\textwidth}{!}{
  \begin{tabular}{lcccccccc}
    \hline
    Ref. & Channel & Final state & $\ell$ & $\tauh$ & AK8 jets & AK8 jet mass & AK4 {\cPqb} jets \\
    \hline
    \cite{Sirunyan:2017acf} & $\PW\PW$, $\PW\PZ$, $\PZ\PZ$ & $\qqbar\qqbar$ & 0 & 0 & $\geq$2 & $2 \, [m_{\PW},m_{\PZ}]$ & \NA \\
    \cite{Sirunyan:2018ivv} & $\PW\PZ$, $\PZ\PZ$ & $\nu\nu\qqbar$ & 0 & 0 & $\geq$1 & $m_\V$ & 0 \\
    \cite{Sirunyan:2018iff} & $\PW\PW$, $\PW\PZ$ & $\ell\nu\qqbar$ & 1 & \NA & $\geq$1 & see Sec.~\ref{sec:selection} & 0 \\
    \cite{Sirunyan:2018hsl} & $\PW\PZ$, $\PZ\PZ$ & $\ell\ell\qqbar$ & $\geq$2 & \NA & $\geq$1 & $m_\V$ & \NA \\
    \cite{Sirunyan:2017jtu} & $\PZ\PZ$ & $\ell\ell\nu\nu$ & $\geq$2 & \NA & \NA & \NA & \NA \\
    
    \cite{Sirunyan:2017wto} & $\PW\PH$, $\PZ\PH$ & $\qqbar\bbbar$ & 0 & 0 & $\geq$2 & $[m_{\PW},m_{\PZ}]$, $m_{\PH}$ & \NA \\
    \cite{Sirunyan:2018qob} & $\PZ\PH$ & $\nu\nu\bbbar$ & 0 & 0 & $\geq$1 & $m_{\PH}$ & 0 \\
    \cite{Sirunyan:2018qob} & $\PW\PH$ & $\ell\nu\bbbar$ & 1 & 0 & $\geq$1 & $m_{\PH}$ & 0 \\
    \cite{Sirunyan:2018qob} & $\PZ\PH$ & $\ell\ell\bbbar$ & $\geq$2 & 0 & $\geq$1 & $m_{\PH}$ & \NA \\
    \cite{Sirunyan:2018fuh} & $\PW\PH$, $\PZ\PH$ & $\qqbar\tau\tau$ & \NA & $\geq$2 & $\geq$1 & $[m_{\PW}, m_{\PZ}]$ & 0 \\
    
    \cite{Sirunyan:2018fuh} & $\PH\PH$ & $\tau\tau\bbbar$ & \NA & $\geq$2 & $\geq$1 & $m_{\PH}$ & 0 \\
    \cite{Sirunyan:2017isc} & $\PH\PH$ & $\bbbar\bbbar$ & \NA & \NA & $\geq$2 & $2 \, m_{\PH}$ & \NA \\
    \cite{Sirunyan:2018qca} & $\PH\PH$ & $\bbbar\bbbar$ & \NA & \NA & $1$ & $m_{\PH}$ & $\geq$2 \\
    
    \cite{Sirunyan:2018mpc} & \multicolumn{2}{c}{$\ell\nu$} & 1 & \NA & \NA & \NA & \NA \\
    \cite{Sirunyan:2018exx} & \multicolumn{2}{c}{$\ell\ell$} & $\geq$2 & \NA & \NA & \NA & \NA \\
    \hline
  \end{tabular}
  }
\label{tab:analyses}
\end{table*}

\section{Systematic uncertainties}\label{sec:syst}

The systematic uncertainties originating from the background estimation in the individual channels are considered uncorrelated, because the backgrounds are determined from statistically independent control regions. The uncertainties arising from the reconstruction and calibration are instead correlated among the different channels. These include the uncertainties in jet energy and resolution, and in the \Pe, \PGm, and $\tauh$ lepton energy, reconstruction, and identification. The uncertainties in the identification of the SM bosons that decay to quarks are dominant in final states containing at least one such decay, and originate from the \mj scale and resolution, the \nsub selection, and the {\cPqb} tagging.
The uncertainties in the \V~tagging extrapolation to large jet \pt and the selection of events in the jet mass window of the Higgs boson are estimated by comparing the results with those obtained using the alternative \HERWIGpp~\cite{Bahr:2008pv} shower model. These uncertainties affect the signal distribution, selection efficiency, and induce event migration effects between search regions.
The uncertainties in the proton-proton inelastic cross section~\cite{Sirunyan:2018nqx}, the integrated luminosity during the 2016 data-taking~\cite{CMS:lumi} and the kinematic acceptance of final-state particles, which affect the signal normalization, are also considered as correlated among channels.
Theoretical uncertainties in the cross section and in the signal geometric acceptance related to the choice of PDFs used in the event generators~\cite{Butterworth:2015oua}, and the uncertainties in the factorization and renormalization scales, are evaluated according to the PDF4LHC recommendations~\cite{Butterworth:2015oua}. The impact of these uncertainties on the signal cross section can be as large as 78\%, depending on the signal mass and the initial state (\qqbar or $\cPg\cPg$). A summary of the main systematic uncertainties is given in Table~\ref{tab:systematics}.

\begin{table*}[!htb]
\centering
\topcaption{Summary of the main systematic uncertainties. The second column reports whether a systematic uncertainty is considered fully correlated or not across different channels. The third column indicates whether the uncertainty affects the yield, the shape of the distributions, or both, or if it induces migration (migr.) effects across search regions. The fourth column reports the smallest and largest effect of the uncertainty, among the yield, the migration, and the signal shape parameters. The symbols ``\SI'', ``\BA'' indicate that the uncertainty affects the signal, the main backgrounds of the analysis, respectively. The treatment of non-dominant backgrounds is often different and not reported here. The symbol ``\FP'' indicates that the parameters are not constrained, or associated with large uncertainties as in the case of multi-dimensional fits. The entries labeled with ``\TE'' are treated differently depending on the interpretation of the exclusion limit, as discussed in Section~\ref{sec:combination}. Uncertainties marked with ``\NA'' are not applicable or are negligible.}
  \resizebox{\textwidth}{!}{
    \begin{tabular}{l ccc cccccccccc cc}
                                      & \rotatebox{90}{Correlation} & \rotatebox{90}{Type} & \rotatebox{90}{Variation} & \rotatebox{90}{$\qqbar\qqbar$~\cite{Sirunyan:2017acf}} & \rotatebox{90}{$\nu\nu\qqbar$~\cite{Sirunyan:2018ivv}} & \rotatebox{90}{$\ell\nu\qqbar$ (2D fit)~\cite{Sirunyan:2018iff}} & \rotatebox{90}{$\ell\ell\qqbar$~\cite{Sirunyan:2018hsl}} & \rotatebox{90}{$\nu\nu\ell\ell$~\cite{Sirunyan:2017jtu}} & \rotatebox{90}{$\qqbar\bbbar$~\cite{Sirunyan:2017wto}} & \rotatebox{90}{$(\nu\nu,\ell\nu,\ell\ell)\bbbar$~\cite{Sirunyan:2018qob}} & \rotatebox{90}{$(\qqbar,\bbbar)\tau\tau$~\cite{Sirunyan:2018fuh}} & \rotatebox{90}{$\bbbar\bbbar$~\cite{Sirunyan:2017isc,Sirunyan:2018qca}} & \rotatebox{90}{$\ell\nu$~\cite{Sirunyan:2018mpc}} & \rotatebox{90}{$\ell\ell$~\cite{Sirunyan:2018exx}} \\
      \hline
      Bkg. modeling                   & \NO  & shape        & \NA      & \FP & \BA & \FP & \BA & \BA & \FP & \BA & \BA & \BA & \BA & \BA \\
      Bkg. normalization              & \NO  & yield        & 2--30\%  & \FP & \BA & \FP & \BA & \BA & \FP & \BA & \BA & \BA & \BA & \BA \\
      Jet energy scale                & \YES & yield, shape & 1--2\%   & \SI & \SI & \SB & \SI & \NA & \SI & \SI & \SI & \SI & \NA & \NA \\
      Jet energy resolution           & \YES & yield, shape & 3--7\%   & \SI & \SI & \SB & \SI & \NA & \SI & \SI & \SI & \SI & \NA & \NA \\
      Jet mass scale                  & \YES & yield, migr. & 1--36\%  & \SI & \SI & \SB & \SI & \NA & \SI & \SI & \SI & \SI & \NA & \NA \\
      Jet mass resolution             & \YES & yield, migr. & 5--25\%  & \SI & \SI & \SB & \SI & \NA & \SI & \SI & \SI & \SI & \NA & \NA \\
      Jet triggers                    & \YES & yield        & 1--15\%  & \SI & \NA & \NA & \NA & \NA & \SI & \NA & \NA & \SI & \NA & \NA \\
      \Pe, \PGm id., iso., trigger    & \YES & yield, shape & 1--3\%   & \NA & \NA & \SI & \SI & \SB & \NA & \SI & \SI & \NA & \SB & \SB \\
      \Pe, \PGm scale and res.        & \YES & yield, shape & 1--6\%   & \NA & \NA & \SI & \SI & \SB & \NA & \SI & \NA & \NA & \SB & \SB \\
      $\tauh$ reco., id., iso.        & \YES & yield        & 6--13\%  & \NA & \SI & \NA & \NA & \NA & \NA & \SI & \SI & \NA & \NA & \NA \\
      $\tauh$ energy scale            & \YES & yield, shape & 1--5\%   & \NA & \NA & \NA & \NA & \NA & \NA & \NA & \SI & \NA & \NA & \NA \\
      $\tauh$ high-\pt extr.          & \YES & yield, shape & 18--30\% & \NA & \NA & \NA & \NA & \NA & \NA & \NA & \SI & \NA & \NA & \NA \\
      \ptmiss scale and res.          & \YES & yield        & 1--2\%   & \NA & \SI & \SI & \NA & \SB & \NA & \SI & \SI & \NA & \SB & \NA \\
      \ptmiss triggers                & \YES & yield        & 1--2\%   & \NA & \SI & \NA & \NA & \NA & \NA & \SI & \SI & \NA & \NA & \NA \\
      \cPqb~quark identification      & \YES & yield, migr. & 1--9\%   & \NA & \SI & \SB & \NA & \NA & \SI & \SI & \SI & \SI & \NA & \NA \\
      $\nsub$ identification          & \YES & yield, migr. & 11--33\% & \SI & \SI & \SB & \SI & \NA & \SI & \NA & \SI & \SI & \NA & \NA \\
      $\V$ tagging high-\pt extr.     & \YES & yield, migr. & 2--40\%  & \SI & \SI & \SB & \SI & \NA & \SI & \NA & \SI & \SI & \NA & \NA \\
      $m_{\PH}$ selection             & \YES & yield        & 6\%      & \NA & \NA & \NA & \NA & \NA & \SI & \SI & \SI & \SI & \NA & \NA \\
      $\Pp\Pp$ cross section          & \YES & yield        & 1--2\%   & \SI & \SI & \NA & \SI & \NA & \SI & \SI & \SI & \SI & \SB & \NA \\
      Luminosity                      & \YES & yield        & 2.5\%    & \SI & \SI & \SI & \SI & \SB & \SI & \SI & \SI & \SI & \SB & \SB \\
      PDF and QCD accept.             & \YES & yield        & 1--2\%   & \SI & \NA & \SI & \SI & \SB & \SI & \SI & \SI & \SI & \NA & \SB \\
      PDF and QCD norm.               & \YES & yield        & 2--78\%  & \TE & \TE & \TE & \TE & \TB & \TE & \TE & \TE & \TE & \TB & \TB \\
      \hline
    \end{tabular}
  }
  \label{tab:systematics}
\end{table*}

\section{Statistical combination}\label{sec:combination}

No significant excess is observed above the SM background expectations.
Upper limits are set at 95\% confidence level (\CL)~\cite{CLS1,CLS2} on the cross section of a heavy resonance, which is rescaled by a signal strength modifier parameter $\mu$ with uniform prior. Systematic uncertainties are represented by nuisance parameters $\theta$, and affect both signal and background expectations~\cite{CMS-NOTE-2011-005}. Systematic uncertainties are considered fully (anti-)correlated when related to a common nuisance parameter, or uncorrelated when different nuisance parameters are used. The prior on these parameters is either flat (represented by the symbol ``\FP'' in Table~\ref{tab:systematics}) or log-normal distributed (identified with ``\SI'', ``\BA'', ``\TE'' in Table~\ref{tab:systematics}). The statistical procedure is based on a likelihood constructed as:
\ifthenelse{\boolean{cms@external}}{
\begin{multline}
\mathcal{L}(\text{data} | \mu, \theta) =\\
 \prod_{c} \prod_{i} \mathcal{P}\left(\text{data} | \mu \, s_{c,i} (\theta) + b_{c,i} (\theta) \right) \, \prod_{j} p_{j} (\tilde{\theta}_{j}|\theta_{j}),
\end{multline}
}{
\begin{equation}
\mathcal{L}(\text{data} | \mu, \theta) = \prod_{c} \prod_{i} \mathcal{P}\left(\text{data} | \mu \, s_{c,i} (\theta) + b_{c,i} (\theta) \right) \, \prod_{j} p_{j} (\tilde{\theta}_{j}|\theta_{j}),
\end{equation}
}
where $\mathcal{P}$ represents the Poisson probability, and $p_{j} (\tilde{\theta}_{j}|\theta_{j})$ is the frequentist \emph{pdf} of the nuisance parameter $\theta_{j}$ and its default value $\tilde{\theta}_{j}$ associated with the $j^\text{th}$ uncertainty. The values $s_{c,i} (\theta)$ and $b_{c,i} (\theta)$ represent the number of signal and background events in the channel $c$ and bin $i$, respectively. The test statistic is based on the profile likelihood ratio $\tilde{q}$:
\begin{equation}
\tilde{q}(\mu) = -2 \log \frac{\mathcal{L}(\text{data} | \mu, \hat{\theta}_{\mu})}{\mathcal{L}(\text{data} | \hat{\mu}, \hat{\theta}_{\hat{\mu}})},
\end{equation}
and the quantities $\hat{\mu}, \hat{\theta}_\mu, \hat{\theta}_{\hat{\mu}}$ are fixed to their best-fit value, and the range of $\mu$ is limited in the $0 \leq \hat{\mu} \leq \mu$ interval.
 The uncertainties that affect the signal normalization (PDFs and factorization and renormalization scales, marked with ``\TE'' in Table~\ref{tab:systematics}) are treated differently depending on how the exclusion is presented. When deriving upper limits on the cross section, these uncertainties are not varied in the fit, but are reported separately as the uncertainty of the theoretical cross sections from the model. When placing limits on the model parameters, these nuisance parameters are fixed at the best-fit values, in the same manner as to the other systematic uncertainties.

The upper limit on $\mu$ is derived from the 95\% \CLs criterion, defined as $\CLs(\mu) = p(\mu)/(1- p(0))$, such that $\CLs(\mu) =  0.05$. The quantities $p(\mu)$ and $1-p(0)$ represent the probabilities to have a value of $\tilde{q}$ equal to, or larger than the observed value in the signal or background ($\mu=0$) hypotheses, respectively, and are derived from analytical functions using the asymptotic approximation~\cite{Asymptotic}. This approximation leads to limits that are up to 30\% stronger in regions with a small number of data events, compared to those obtained from the Monte Carlo generation of pseudo-experiments~\cite{CMS-NOTE-2011-005}.

\section{Results and interpretation}\label{sec:results}

The data are in agreement with the expected background from SM processes.
The largest deviation from the expected limit is observed in the \PVpr~model~A at a mass of 1.3\TeV, with a local significance of 2.7 standard deviations, corresponding to a global significance of 1.6 standard deviations.
The local significances are obtained in the asymptotic approximation~\cite{Asymptotic}, and the global significances are evaluated using the trial factors method~\cite{Gross:2010qma}.

\begin{figure}[!hbt]\centering
  \includegraphics[width=0.495\textwidth]{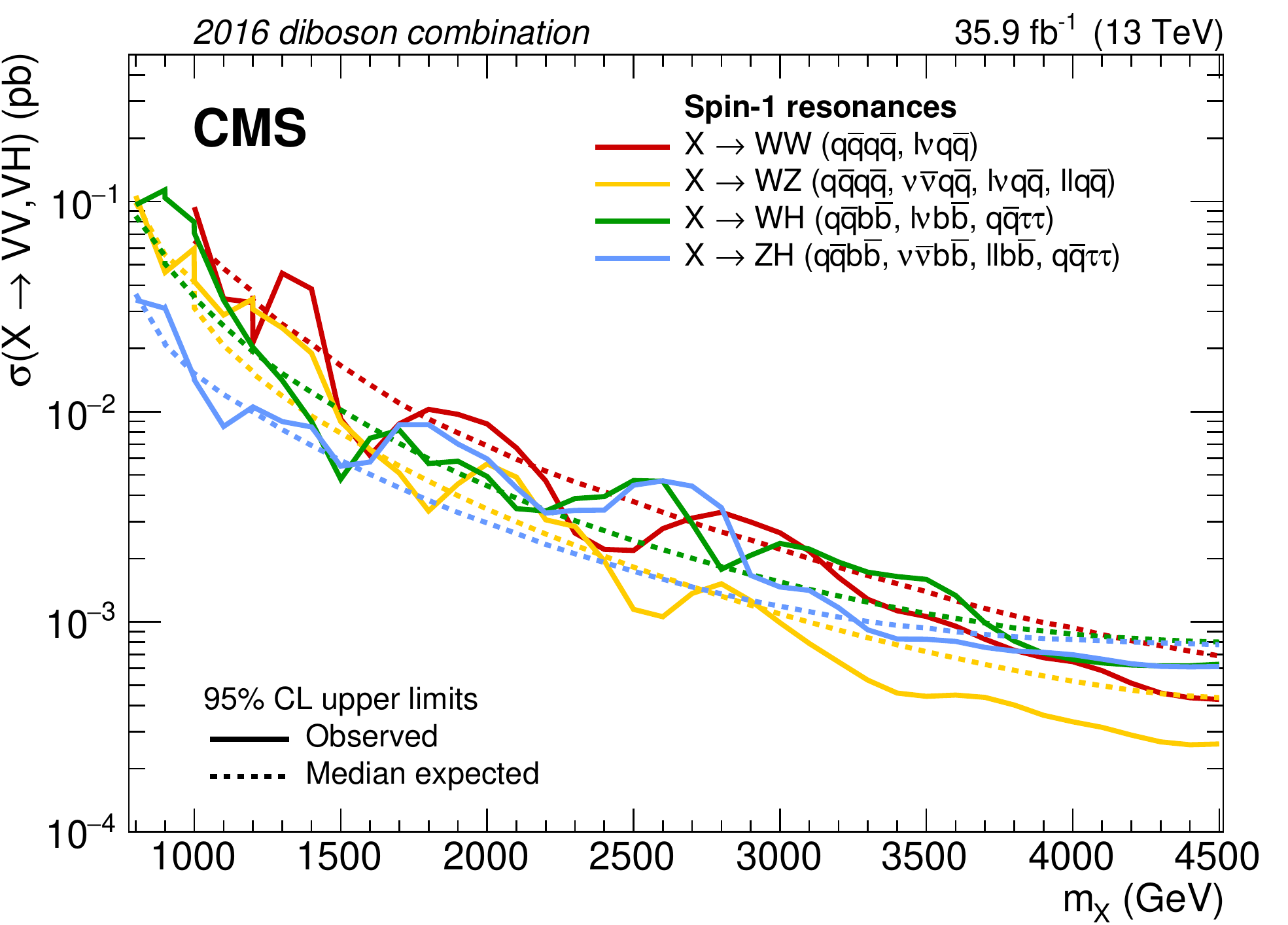}
  \includegraphics[width=0.495\textwidth]{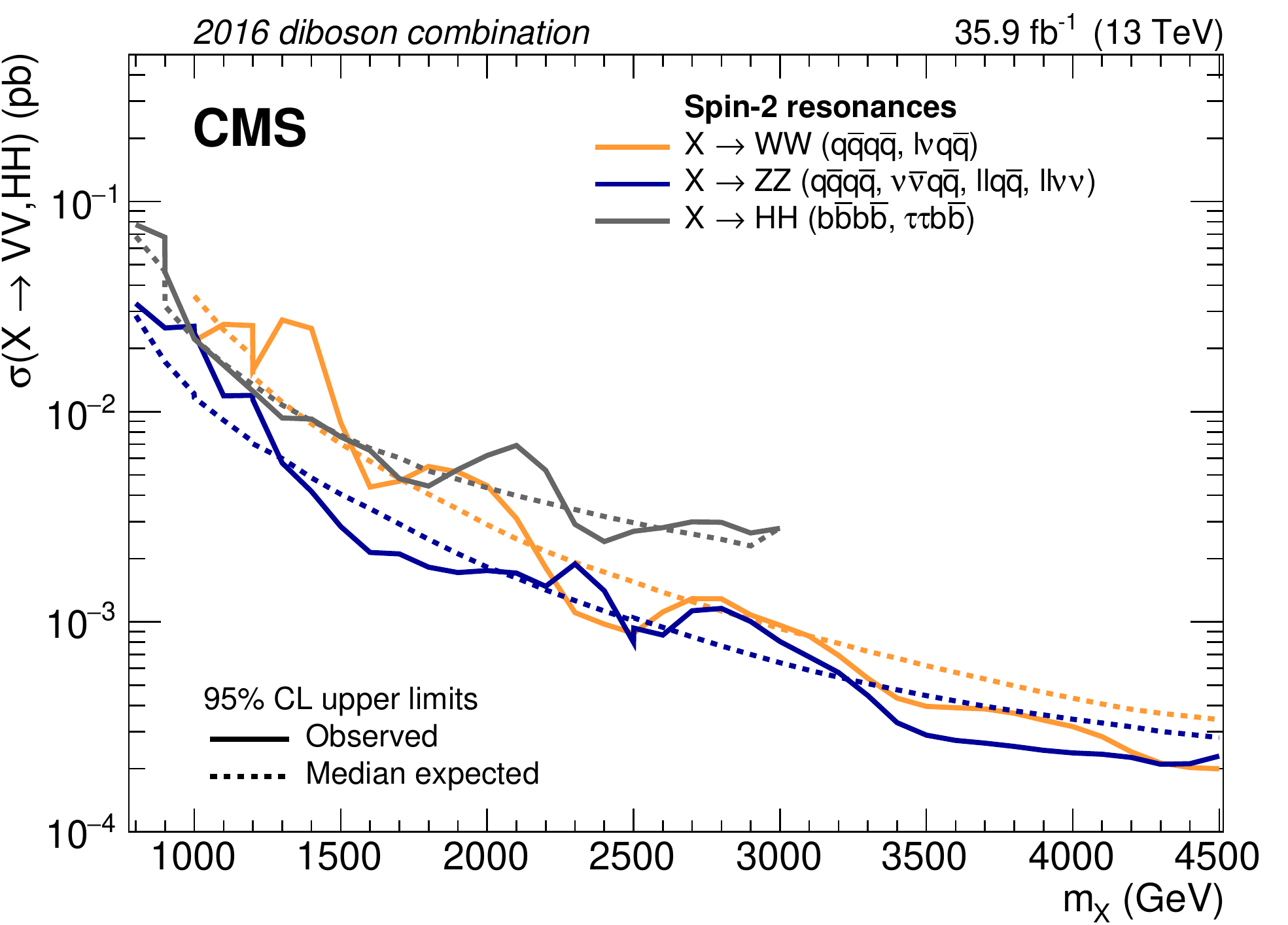}
  \caption{Observed and expected 95\% \CL upper limits on the product of the cross section and branching fraction of a spin-1 (\cmsLeft) or spin-2 resonance (\cmsRight) decaying to a pair of SM bosons.}
    \label{fig:diboson}
\end{figure}

The exclusion limit on the cross section of each diboson channel ($\PW\PW$, $\PW\PZ$, $\PZ\PZ$, $\PW\PH$, $\PZ\PH$, $\PH\PH$) is depicted in Fig.~\ref{fig:diboson} for the combination of all contributing channels, reported in Table~\ref{tab:analyses}, according to the spin of the new resonance. The generated signal can be either a spin-1 heavy vector (\PWpr or \PZpr, as in the HVT model) or a spin-2 boson (as in the bulk graviton model). In fact, the spin and polarization of the heavy resonance does affect the final state, signal acceptance, and selection efficiencies. The exclusion limits are not presented above 4.5\TeV, since at larger masses the background estimation procedure used in diboson analyses becomes less reliable because of the lack of events in data.

\begin{figure}[!hbt]\centering
  \includegraphics[width=0.495\textwidth]{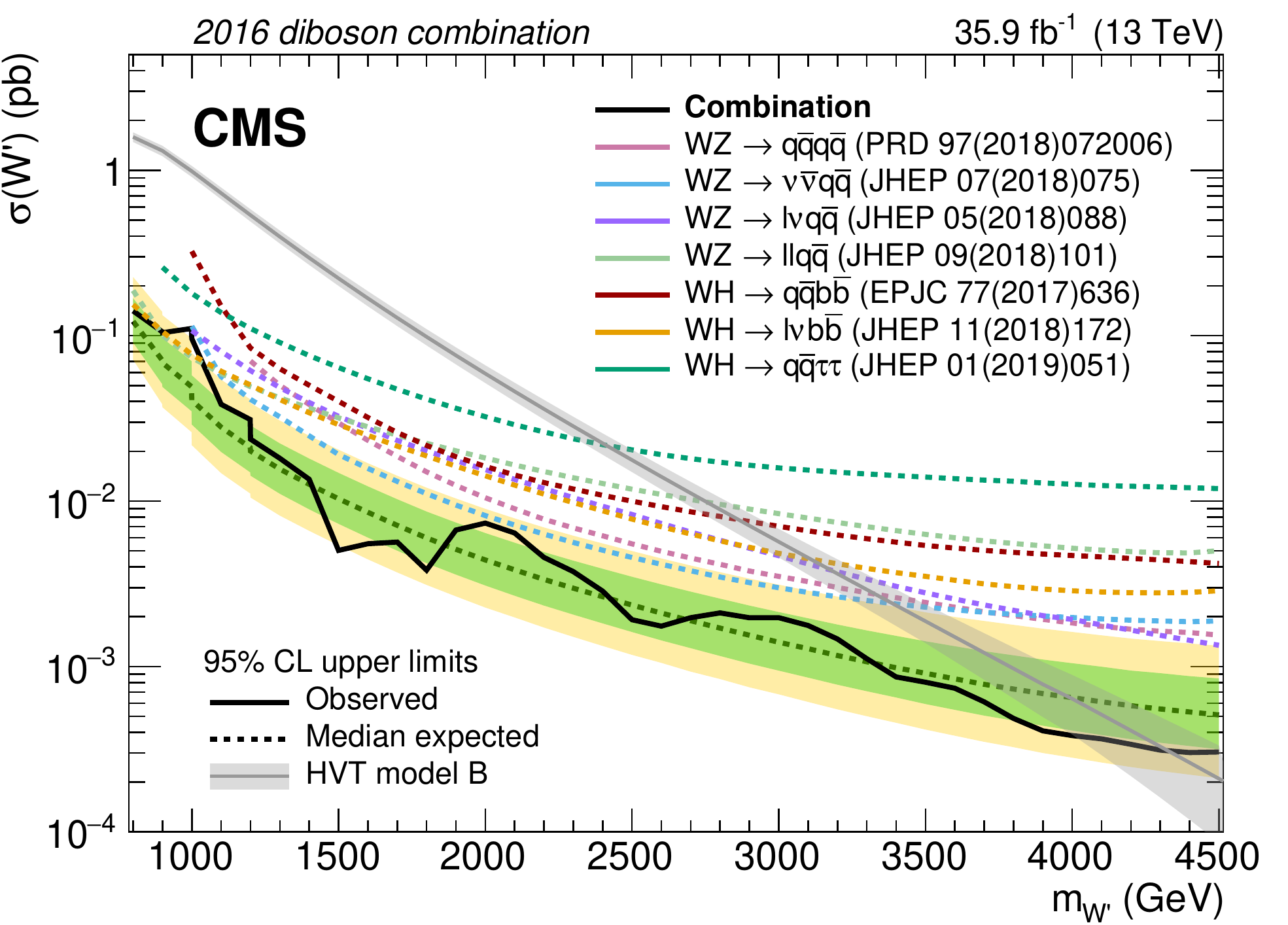}
  \includegraphics[width=0.495\textwidth]{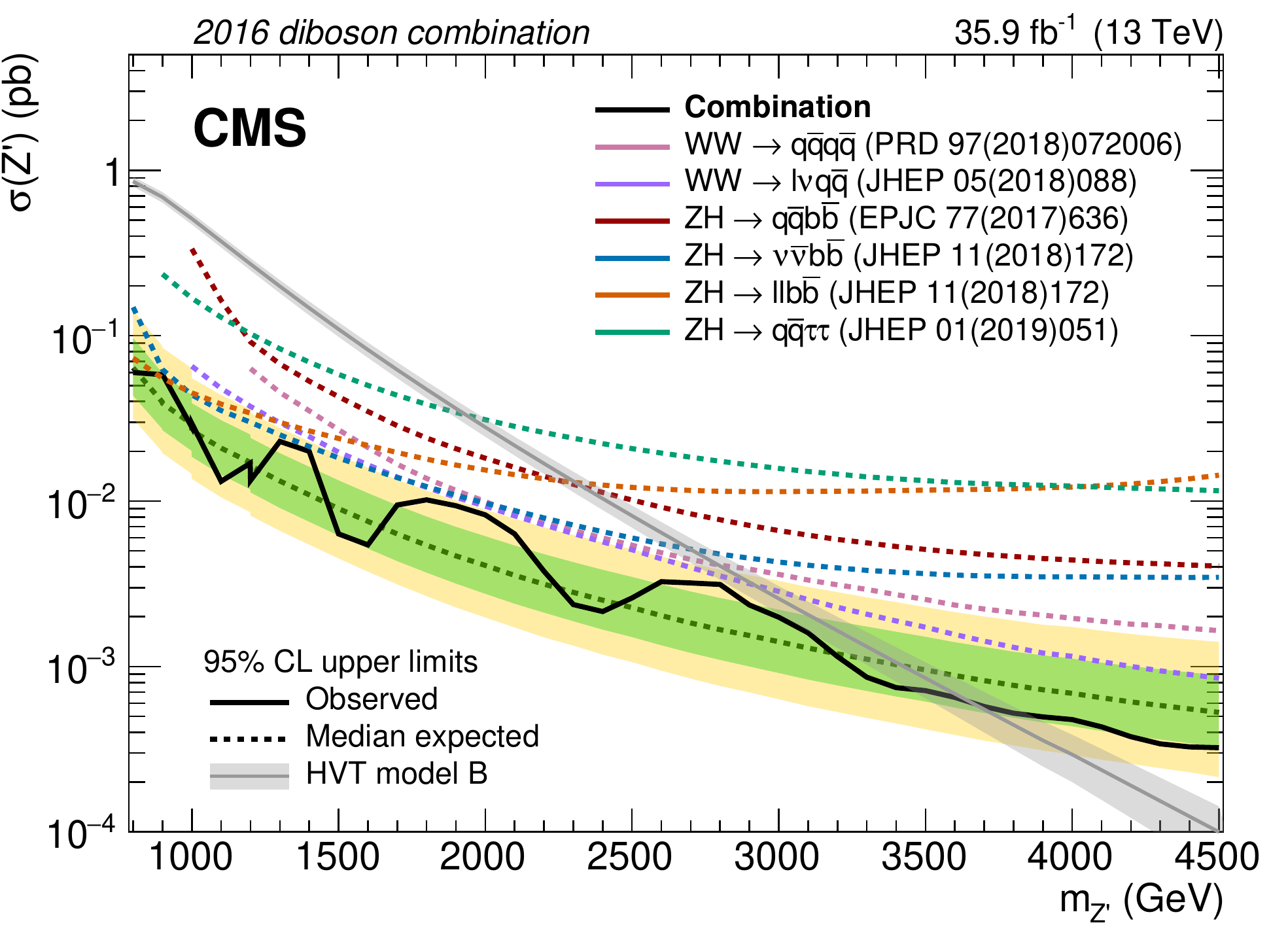}
  \caption{Observed and expected 95\% \CL upper limits on the \PWpr (\cmsLeft) and \PZpr cross section (\cmsRight) as a function of the \PWpr and \PZpr resonance mass. The inner green and outer yellow bands represent the ${\pm}1$ and ${\pm}2$ standard deviation variations on the expected limits of the statistical combination of the $\V\V$ and $\V\PH$ channels considered. The expected limits in individual channels are represented by the colored dashed lines. The solid curves surrounded by the shaded areas show the cross sections predicted by the HVT model~B and their uncertainties.}
    \label{fig:singlet}
\end{figure}

The combined exclusion limits for the spin-1 singlet hypotheses (\PWpr or \PZpr) in the HVT model~B framework, where the branching fractions to SM bosons are dominant, are shown in Fig.~\ref{fig:singlet}. In this scenario, the contribution of the dilepton channels is negligible because their branching fraction is of the order of a few permil. In the $\V\V$ analyses, the $\PVpr\to\V\PH$ signal is added to the $\PVpr\to\V\V$ signal because the fraction of \PH jets passing the $m_{\PW}$ and $m_{\PZ}$ selections is not negligible and can contribute up to 30\% to the total signal yield. The predictions of the HVT model~B are superimposed on the exclusion limits, showing that a \PWpr boson of mass below 4.3\TeV, and a \PZpr boson with mass below 3.7\TeV are excluded at 95\% \CL.

\begin{figure}[!hbt]\centering
  \includegraphics[width=0.495\textwidth]{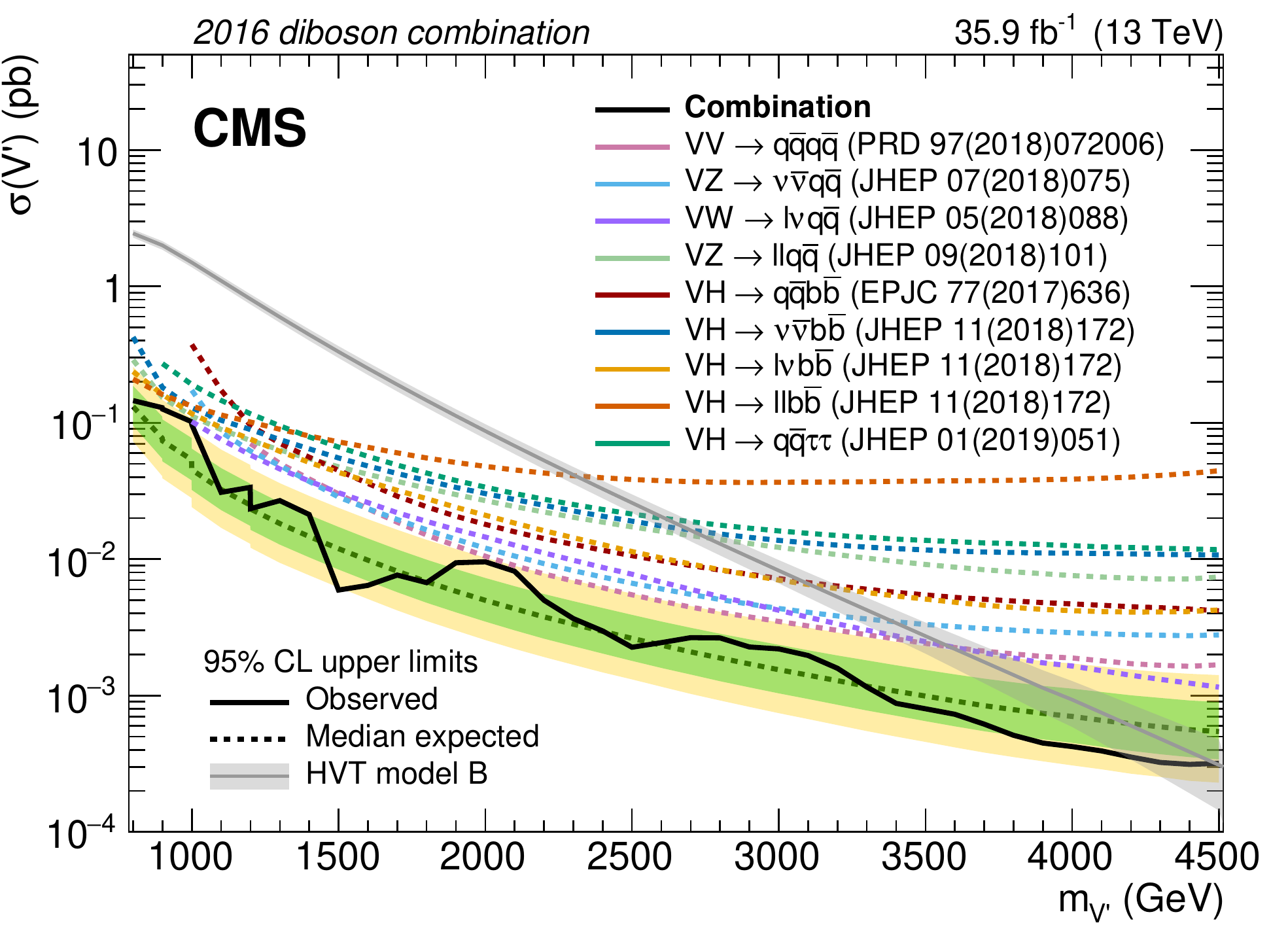}
  \includegraphics[width=0.495\textwidth]{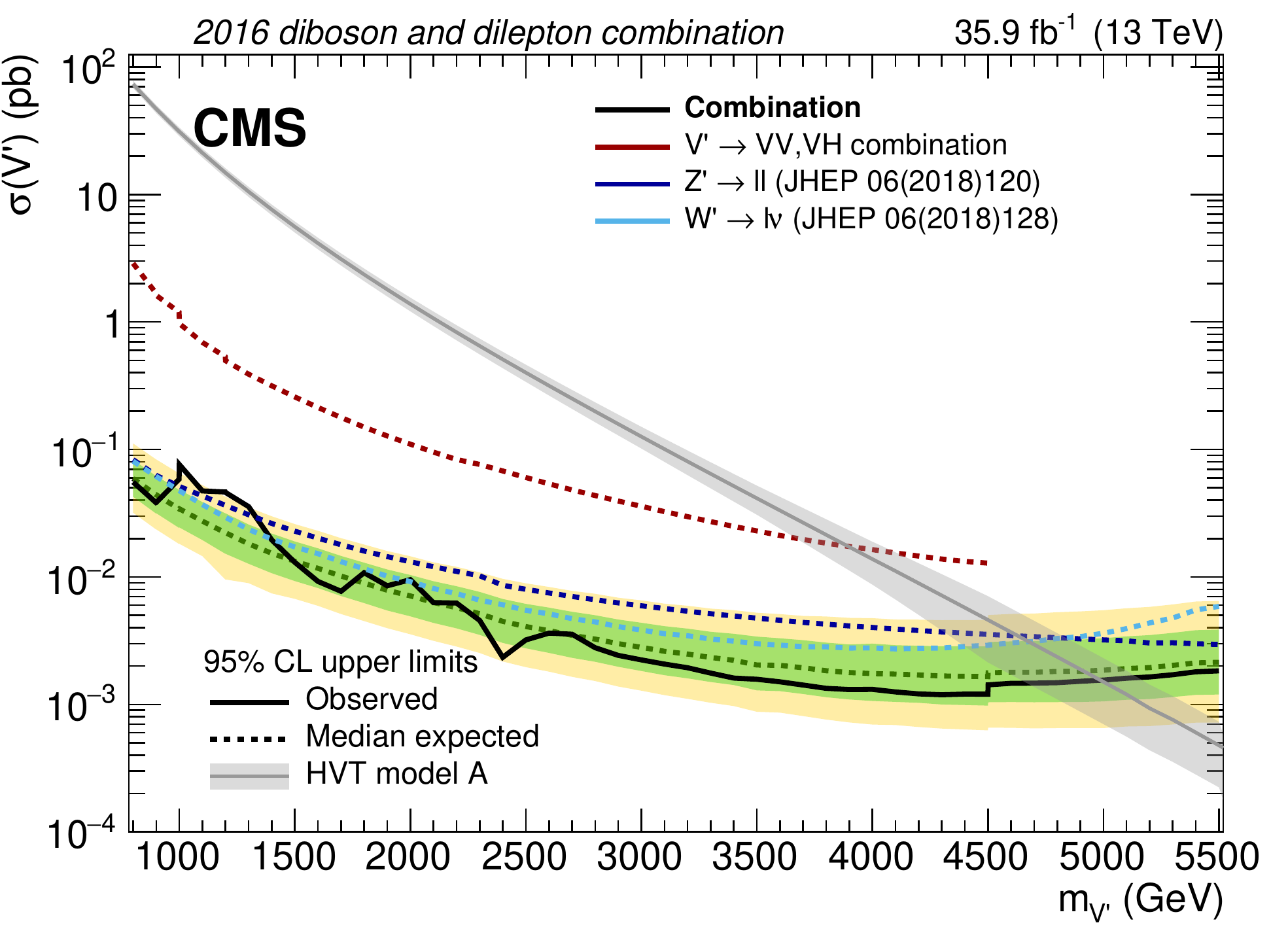}
  \caption{Observed and expected 95\% \CL upper limits on cross sections as a function of the HVT triplet mass for the combination of all channels in the HVT model~B (\cmsLeft) and model~A (\cmsRight). The inner green and outer yellow bands represent the ${\pm}1$ and ${\pm}2$ standard deviation variations on the expected limit. The solid curves surrounded by the shaded areas show the cross sections predicted by HVT models~A and B and their uncertainties.}
    \label{fig:triplet}
\end{figure}

The HVT hypothesis is tested in Fig.~\ref{fig:triplet} by combining all diboson channels, showing that a mass-degenerate state with mass below 4.5\TeV can be excluded in HVT model~B, and extending the exclusion with respect to the best individual channel by approximately 700\GeV~\cite{Sirunyan:2017acf}.
This result significantly improves the previous $\sqrt{s}=8$ and $13\TeV$ CMS combination~\cite{Sirunyan:2017nrt}, which excluded a triplet of heavy resonances with masses up to 2.4\TeV in the same model.
The dilepton resonances provide the most stringent results within the HVT model~A framework, and are combined with the diboson searches in Fig.~\ref{fig:triplet}. A heavy triplet of \PVpr~resonances is excluded up to a mass of 5.0\TeV.

\begin{figure}[!hb]\centering
  \includegraphics[width=\cmsFigWidth]{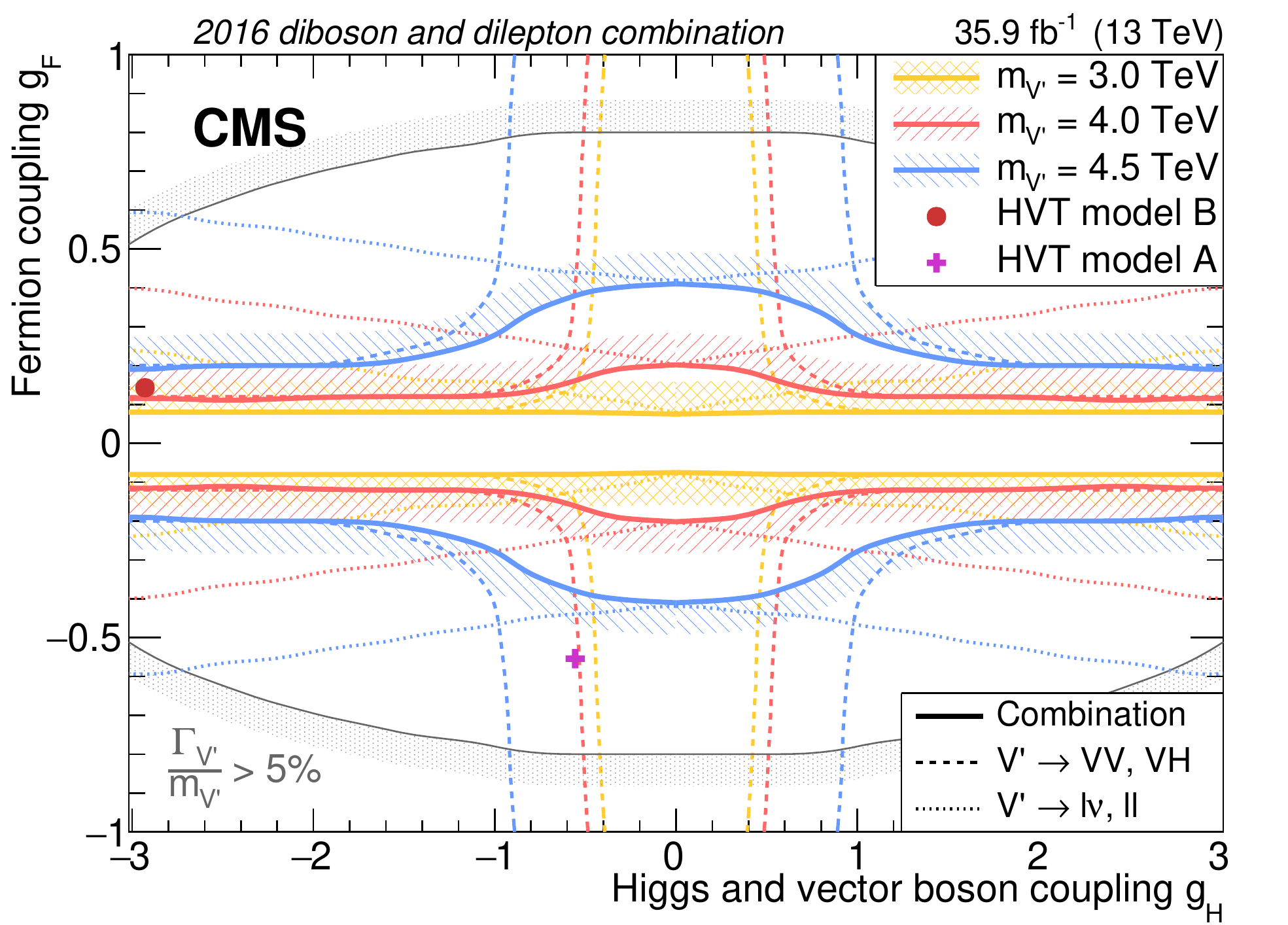}
  \caption{Observed exclusion limits on the couplings of heavy vector resonances to fermions and SM vector bosons and the Higgs boson for the statistical combination (solid lines) of the dilepton (dotted lines) and diboson channels (dashed lines). Three resonance masses hypotheses (3.0, 4.0, and 4.5\TeV) are considered. The hatched bands indicate the regions excluded. The areas bounded by the thin gray contour lines correspond to regions where the resonance widths ($\Gamma_{\PVpr}$) are predicted to be larger than the average experimental resolution (5\%).}
    \label{fig:hvt}
\end{figure}

The exclusion limits on the resonance cross sections shown in Fig.~\ref{fig:triplet} are also interpreted as limits in the $\left[\gH, \gF\right]$ plane of the HVT parameters. The excluded region of parameter space for narrow resonances obtained from the combination of all the channels is shown in Fig.~\ref{fig:hvt}. The dilepton and diboson searches constrain different regions of the parameter space, as the dilepton searches can probe the region where the coupling to the SM bosons approaches zero. In the triplet interpretation, the ratio of the \PWpr to \PZpr cross sections is assumed to be determined by the ratio of the partonic luminosities, and to depend only weakly on the model parameters. The fraction of the parameter space where the natural width of the resonances is larger than the average experimental resolution of 5\%, and the narrow-width approximation is thus invalid, is also indicated in Fig.~\ref{fig:hvt}.

\begin{figure}[!hb]\centering
  \includegraphics[width=\cmsFigWidth]{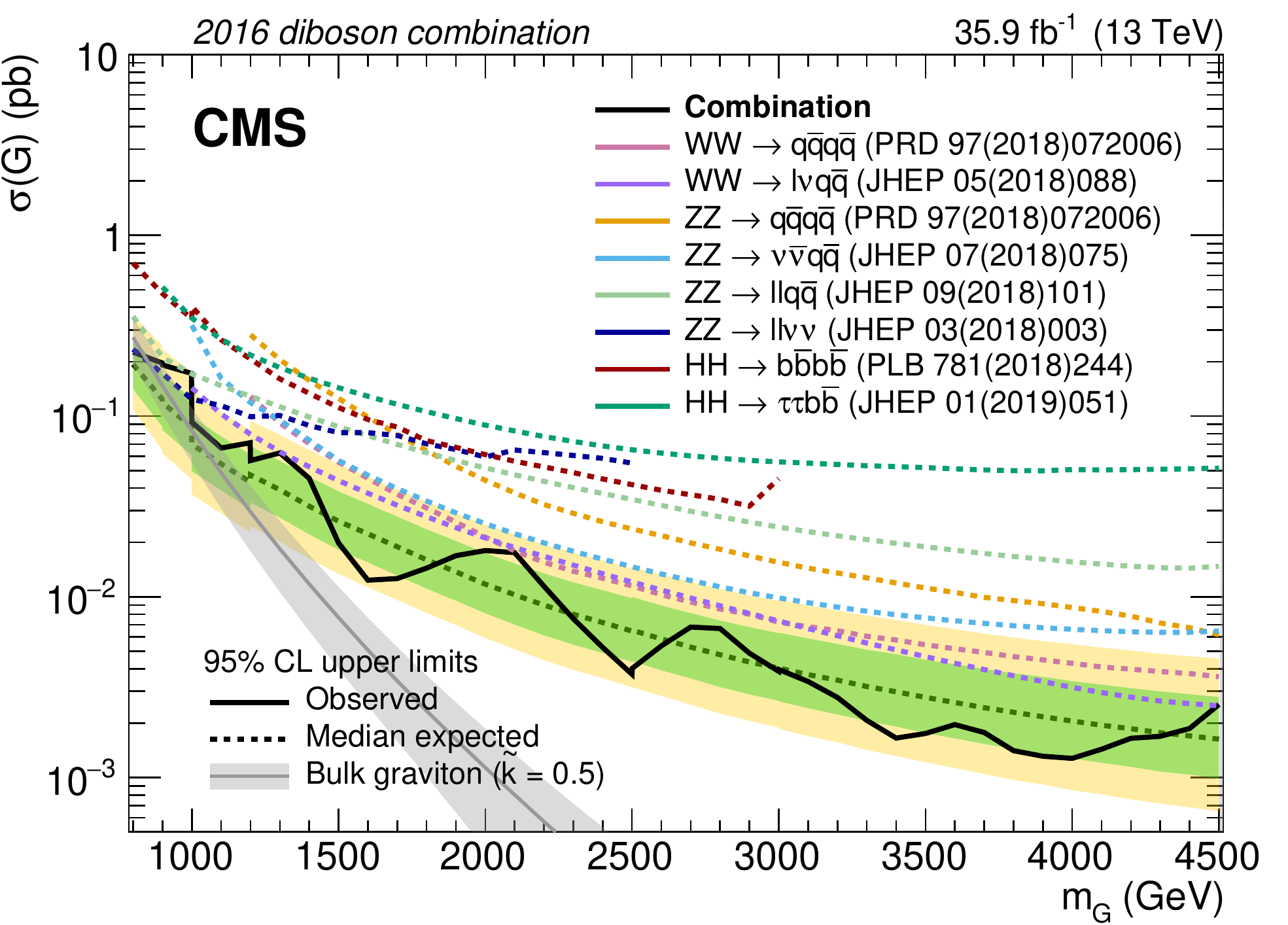}
  \caption{Observed and expected 95\% \CL upper limit on the cross section of the spin-2 bulk graviton as a function of its mass for the statistical combination of the $\PW\PW$, $\PZ\PZ$, and $\PH\PH$ channels. The inner green and outer yellow bands represent the ${\pm}1$ and ${\pm}2$ standard deviation variations on the expected limit. The solid curve and its shaded area represent the cross section derived with the parameter $\ktilde=0.5$ and the associated uncertainty.}
    \label{fig:G}
\end{figure}

In the spin-2 bulk graviton model, the $\PW\PW$, $\PZ\PZ$, and $\PH\PH$ channels are combined, setting upper limits of up to 1.1\unit{fb} on the cross section of a graviton with mass up to 4.5\TeV. In the $\ktilde=0.5$ scenario, a graviton with a mass smaller than 850\GeV is excluded at 95\% \CL, as shown in Fig.~\ref{fig:G}. Larger \ktilde values increase the production cross sections, but also the graviton natural width, which can be comparable or larger than the experimental resolution. In these cases, the narrow-width approximation is no longer valid.

These results represent the most stringent limits on heavy vector models set by CMS and are comparable at large $\PVpr$ mass to the limits obtained the ATLAS in the combination of similar channels~\cite{Aaboud:2018bun}. At lower masses, the ATLAS combination excludes smaller cross sections because of the inclusion of final states with three or more leptons. The exclusion limits in the bulk graviton model are not directly comparable because of the large $\ktilde=1.0$ parameter adopted by ATLAS, which increases the cross section but also implies a non-negligible natural width.

\section{Summary}\label{sec:conclusion}

A statistical combination of searches for heavy resonances decaying into pairs of vector bosons, a vector boson and a Higgs boson, two Higgs bosons, or pairs of leptons, has been presented. The results are based on data collected by the CMS experiment at $\sqrt{s}=13\TeV$ during 2016 corresponding to an integrated luminosity of 35.9\fbinv. In models with warped extra dimensions, upper limits of up to 1.1\unit{fb} are set at 95\% confidence level on the production cross section of the spin-2 bulk graviton. For models with a triplet of narrow spin-1 resonances, heavy vector bosons with masses below 5.0 and 4.5\TeV are excluded at 95\% confidence level in models where the \PWpr and \PZpr bosons couple predominantly to fermions and bosons, respectively. In the latter, the statistical combination extends the exclusion limit by 700\GeV as compared to the best individual channel.

\begin{acknowledgments}
We congratulate our colleagues in the CERN accelerator departments for the excellent performance of the LHC and thank the technical and administrative staffs at CERN and at other CMS institutes for their contributions to the success of the CMS effort. In addition, we gratefully acknowledge the computing centers and personnel of the Worldwide LHC Computing Grid for delivering so effectively the computing infrastructure essential to our analyses. Finally, we acknowledge the enduring support for the construction and operation of the LHC and the CMS detector provided by the following funding agencies: BMBWF and FWF (Austria); FNRS and FWO (Belgium); CNPq, CAPES, FAPERJ, FAPERGS, and FAPESP (Brazil); MES (Bulgaria); CERN; CAS, MoST, and NSFC (China); COLCIENCIAS (Colombia); MSES and CSF (Croatia); RPF (Cyprus); SENESCYT (Ecuador); MoER, ERC IUT, PUT and ERDF (Estonia); Academy of Finland, MEC, and HIP (Finland); CEA and CNRS/IN2P3 (France); BMBF, DFG, and HGF (Germany); GSRT (Greece); NKFIA (Hungary); DAE and DST (India); IPM (Iran); SFI (Ireland); INFN (Italy); MSIP and NRF (Republic of Korea); MES (Latvia); LAS (Lithuania); MOE and UM (Malaysia); BUAP, CINVESTAV, CONACYT, LNS, SEP, and UASLP-FAI (Mexico); MOS (Montenegro); MBIE (New Zealand); PAEC (Pakistan); MSHE and NSC (Poland); FCT (Portugal); JINR (Dubna); MON, RosAtom, RAS, RFBR, and NRC KI (Russia); MESTD (Serbia); SEIDI, CPAN, PCTI, and FEDER (Spain); MOSTR (Sri Lanka); Swiss Funding Agencies (Switzerland); MST (Taipei); ThEPCenter, IPST, STAR, and NSTDA (Thailand); TUBITAK and TAEK (Turkey); NASU and SFFR (Ukraine); STFC (United Kingdom); DOE and NSF (USA).

\hyphenation{Rachada-pisek} Individuals have received support from the Marie-Curie program and the European Research Council and Horizon 2020 Grant, contract Nos.\ 675440, 752730, and 765710 (European Union); the Leventis Foundation; the A.P.\ Sloan Foundation; the Alexander von Humboldt Foundation; the Belgian Federal Science Policy Office; the Fonds pour la Formation \`a la Recherche dans l'Industrie et dans l'Agriculture (FRIA-Belgium); the Agentschap voor Innovatie door Wetenschap en Technologie (IWT-Belgium); the F.R.S.-FNRS and FWO (Belgium) under the ``Excellence of Science -- EOS" -- be.h project n.\ 30820817; the Beijing Municipal Science \& Technology Commission, No. Z181100004218003; the Ministry of Education, Youth and Sports (MEYS) of the Czech Republic; the Lend\"ulet (``Momentum") Program and the J\'anos Bolyai Research Scholarship of the Hungarian Academy of Sciences, the New National Excellence Program \'UNKP, the NKFIA research grants 123842, 123959, 124845, 124850, 125105, 128713, 128786, and 129058 (Hungary); the Council of Science and Industrial Research, India; the HOMING PLUS program of the Foundation for Polish Science, cofinanced from European Union, Regional Development Fund, the Mobility Plus program of the Ministry of Science and Higher Education, the National Science Center (Poland), contracts Harmonia 2014/14/M/ST2/00428, Opus 2014/13/B/ST2/02543, 2014/15/B/ST2/03998, and 2015/19/B/ST2/02861, Sonata-bis 2012/07/E/ST2/01406; the National Priorities Research Program by Qatar National Research Fund; the Ministry of Science and Education, grant no. 3.2989.2017 (Russia); the Programa Estatal de Fomento de la Investigaci{\'o}n Cient{\'i}fica y T{\'e}cnica de Excelencia Mar\'{\i}a de Maeztu, grant MDM-2015-0509 and the Programa Severo Ochoa del Principado de Asturias; the Thalis and Aristeia programs cofinanced by EU-ESF and the Greek NSRF; the Rachadapisek Sompot Fund for Postdoctoral Fellowship, Chulalongkorn University and the Chulalongkorn Academic into Its 2nd Century Project Advancement Project (Thailand); the Welch Foundation, contract C-1845; and the Weston Havens Foundation (USA).
\end{acknowledgments}

\bibliography{auto_generated}

\cleardoublepage \appendix\section{The CMS Collaboration \label{app:collab}}\begin{sloppypar}\hyphenpenalty=5000\widowpenalty=500\clubpenalty=5000\vskip\cmsinstskip
\textbf{Yerevan Physics Institute, Yerevan, Armenia}\\*[0pt]
A.M.~Sirunyan$^{\textrm{\dag}}$, A.~Tumasyan
\vskip\cmsinstskip
\textbf{Institut für Hochenergiephysik, Wien, Austria}\\*[0pt]
W.~Adam, F.~Ambrogi, T.~Bergauer, J.~Brandstetter, M.~Dragicevic, J.~Erö, A.~Escalante~Del~Valle, M.~Flechl, R.~Frühwirth\cmsAuthorMark{1}, M.~Jeitler\cmsAuthorMark{1}, N.~Krammer, I.~Krätschmer, D.~Liko, T.~Madlener, I.~Mikulec, N.~Rad, J.~Schieck\cmsAuthorMark{1}, R.~Schöfbeck, M.~Spanring, D.~Spitzbart, W.~Waltenberger, J.~Wittmann, C.-E.~Wulz\cmsAuthorMark{1}, M.~Zarucki
% ...
\vskip\cmsinstskip
\dag: Deceased\\
1:  Also at Vienna University of Technology, Vienna, Austria\\
% ...
\end{sloppypar}
\end{document}